\documentclass[conference]{IEEEtran}
\usepackage{amsfonts}
\IEEEoverridecommandlockouts

\ifCLASSINFOpdf
\else
\fi
\usepackage{pdfpages}
\usepackage{subfigure}
\usepackage{epsfig}
\usepackage{graphicx}
\usepackage{psfig}
\usepackage{epsf}
\usepackage[cmex10]{amsmath}
\usepackage{booktabs}
\usepackage{fancyhdr}
\usepackage{amsmath}
\usepackage{caption2}   

\begin{document}
\title{Data-and-Knowledge Dual-Driven Automatic Modulation Recognition for Wireless Communication Networks}
\author{Rui Ding$^{ \S}$, Hao Zhang$^{ \S}$, Fuhui Zhou$^{ \S }$, Qihui Wu$^{ \S }$, Zhu Han$^\dagger$ \\
$^{\S}$Nanjing University of Aeronautics and Astronautics, China, $^\dagger$University of Houston, USA\\

Email: \emph{\{rui\_ding, haozhangcn@nuaa.edu.cn, zhoufuhui@ieee.org, wuqihui2014@sina.com, and zhan2@uh.edu\}}
\thanks{
This work was supported by National Key R\&D Program of China under Grant 2020YFB1807602, the National Natural Science Foundation of China under Grant 62071223, Grant 62031012 and Grant 61931011, Young Elite Scientist Sponsorship Program by CAST.
}}
\maketitle
\begin{abstract}
Automatic modulation classification is of crucial importance in wireless communication networks. Deep learning based automatic modulation classification schemes have attracted extensive attention due to the superior
accuracy. However, the data-driven method relies on a large amount of training samples and the classification accuracy is poor in the low signal-to-noise radio (SNR). In order to tackle these problems, a novel data-and-knowledge dual-driven automatic modulation classification scheme based on radio frequency machine learning is proposed by exploiting the attribute features of different modulations. The visual model is utilized to extract visual features. The attribute learning model is used to learn the attribute semantic representations. The transformation model is proposed to convert the attribute representation into the visual space. Extensive simulation results demonstrate that our proposed automatic modulation classification scheme can achieve better performance than the benchmark schemes in terms of the
classification accuracy, especially in the low SNR. Moreover, the confusion among high-order modulations is reduced by using our proposed scheme compared with other traditional schemes.
\end{abstract}
\IEEEpeerreviewmaketitle
\section{Introduction}
\IEEEPARstart 
Artificial intelligence and intelligent communication are indispensable parts in wireless communication systems. As an indispensable intelligent communication technology in wireless communication systems, automatic modulation classification (AMC) has been widely used in various applications. In military applications, AMC helps to recover the transmitted information and generate interference signals with matching modulation \cite{military}. In civilian applications, AMC is able to determine the appropriate demodulation method to realize the correct recovery of transmitted information \cite{civil}.

The existing automatic modulation classification (AMC) schemes can be mainly classified into two categories, namely, model-driven AMC and data-driven AMC. The model-driven schemes mainly include likelihood-based (LB) schemes \cite{hameed2009likelihood} and feature-based (FB) schemes \cite{9463441}. The LB classifier treats AMC as a hypothesis testing problem. Different test statistics are constructed firstly. Then, the likelihood functions are calculated under the modulation hypothesis by using the constructed test statistics. Lastly, these functions are compared to make the final decision \cite{hameed2009likelihood}. The FB schemes aim to extract unique characteristics of different types of signals and have
robust performance with low implementation complexity \cite{7776899}.
In contrast, data-driven methods, such as support vector machine (SVM) \cite{6183517}, and logistic regression \cite{7776899}, etc., perform modulation classification by learning the difference among data distributions. 
Moreover, the data-driven deep learning utilize the neural networks to extract the visual features automatically from the original data, such as I/Q samples. The related works are classified as follows.

\emph{Model-driven methods}: The authors in \cite{823550} used the maximum-likelihood method to recognize digital amplitude-phase modulations. It was shown that the maximum-likelihood classifier is capable of classifying any finite set of distinctive constellations with the zero error rate when the number of available symbols goes to infinity. However, the maximum-likelihood method suffers from high computational complexity. With the FB schemes, the authors in \cite{8954676} utilized the high-order statistical features to realize the modulation classification. It was shown that the FB method can achieve good performance with low computational complexity.



\emph{Data-driven methods}: A long short-term memory (LSTM) based AMC algorithm was proposed in \cite{9170506}. LSTM learns the dependency relationship between the current element and the elements before-after through the gating structure.
However, its recurrent structure results in high computational complexity. Meanwhile, spatial correlation features are ignored in this scheme.
The authors in \cite{9220797} utilized the CNN-LSTM to efficiently explore the feature interaction and the spatial-temporal properties of raw complex temporal signals. However, the increase of the depth of the network can cause the gradient vanishing and over-fitting problems.
A deep residual network (ResNet) was proposed in \cite{he2016deep} by using residual learning with skip connection for image classifications, which alleviates the over-fitting problem when training deep networks. Meanwhile, it is able to learn discriminative features for achieving a better performance. However, the complex architecture of deep network needs a lot of computing resources and takes a long time to train. This drawback makes it unrealistic in the practical scenarios since the real-time performance is of crucial importance in practical applications.
In \cite{8978670}, the authors designed a lightweight model with smaller model sizes. High computation efficiency was achieved by using model compression. The simulation results demonstrated that the lightweight model reduces the training time significantly with negligible loss in classification accuracy.

All the data-driven AMC schemes mentioned in \cite{9170506}-\cite{8978670} require a large amount of training samples, which are difficult to be obtained in practical communication scenarios. Meanwhile, pure data-driven schemes cannot satisfy the classification performance requirements under the dynamically changing communication scenarios. In particular, the classification performance achieved by using those methods is very poor in low SNRs.
Recently, radio frequency machine learning (RFML) has been proposed \cite{wong2020rfml},\cite{wong2020rfml2} and envisioned to be promising to tackle these problems. It exploits expert knowledge to achieve superior performance. Motivated by RFML, a novel data and knowledge dual-driven AMC scheme is proposed in this paper. Our main contributions are as follows. 
\begin{itemize}
\item It is the first time that semantic information, e.g., class attributes, is exploited to decrease the required number of training samples, which is of crucial importance in the practical complex and dynamic wireless networks.
\item An improved residual network is proposed for attribute learning.
\item A novel data and knowledge dual-driven framework for AMC is proposed to construct the classifier to learn the muti-dimensional representations of different modulations from I/Q signals.
\item Simulation results demonstrate that our proposed scheme has a better classification performance compared with other DL-based AMC schemes, especially in the low SNR. Moreover, the confusion between 16QAM and 64QAM is reduced significantly.
\end{itemize}

The remainder of this paper is organized as follows. The preliminary is presented in Section \ref{Preliminaries}. Section \ref{Proposed AMC Scheme} presents our proposed scheme. Section \ref{Simulation Results} presents the simulation results. Finally, the paper concludes with Section \ref{Conclusion}.

\section{Preliminaries}\label{Preliminaries}
\subsection{Problem Statement}
The modulation classification can be identified as a $K$-class hypothesis test, where $K$ denotes the number of modulations. The received signal under the $k$th modulation hypothesis ${H}_{k}$ is given as
\begin{align}\label{27}\
{{H}_{k}}:{{x}_{k}}(n)={{s}_{k}}(n)+{{\omega }_{k}}(n),n=1,2,...,N,
\end{align}
where ${s}_{k}(n)$ and ${x}_{k}(n)$ are the transmitted signal and the received signal, respectively. $N$ is the number of signal symbols. ${{\omega }_{k}}(n)$ is the additive white Gaussian noise with mean being zero and variance being ${{\sigma }^{2}}$.

The in-phase and quadrature (I/Q) parts of the  received signal are both utilized. These two parts usually obey an identical independent
distribution, which can be input into the neural network without normalization \cite{o2018over}.
The I/Q signal samples can be expressed as a vector by turning the received signal ${x}_{k}(n)$ into the vector $\mathbf{x}_{k}$, given as
\begin{subequations}
\begin{align}
{\mathbf{x}_{k}}&={\mathbf{I}_{k}}+{\mathbf{Q}_{k}} \\
& =\Re({\mathbf{x}_{k}})+j\Im({\mathbf{x}_{k}}),
\end{align}
\end{subequations}
where $\mathbf{I}_{k}$ and $\mathbf{Q}_{k}$ represent the in-phase part and the quadrature part of the received signal, respectively, and $j=\sqrt{-1}$. $\Re(\cdot)$ and $\Im(\cdot)$ represent the operators of the real and imaginary parts of the signal, respectively. The raw data $\mathbf{x}_{k}$ can be specifically expressed as the form of matrix, given as
\begin{align}
{\mathbf{x}_{k}}=\left[
\begin{matrix}
   {\Re}\left[ x(1),x(2),...,x(N)\right] \\
   {\Im}\left[ x(1),x(2),...,x(N)\right] \\
\end{matrix} \right].
\end{align}

\subsection{Traditional AMC Methods}
The DL-based schemes such as CNN and RNN are utilized to extract features from the raw data. Then, a fully
connected (FC) layer is utilized to integrate the information and carry out the conversion of feature dimension. The feature learning can be expressed as a process that the raw data $\mathbf{x}_{k} \in {\mathbb{R}^{N\times 2}}$ is mapped into a $L$-dimensional vector $\mathbf{y}$, given as
\begin{align}
f:{\mathbf{x}_{k}}\in {\mathbb{R}^{N\times 2}}\to \mathbf{y}\in {\mathbb{R}^{L}},
\end{align}
where mapping function $f$ represents the feature learning model with the fully connected layer, and $\mathbf{y}$ represents the feature vector output from the FC layer.

\begin{figure*}[!t]
\centering
\includegraphics[width=7in]{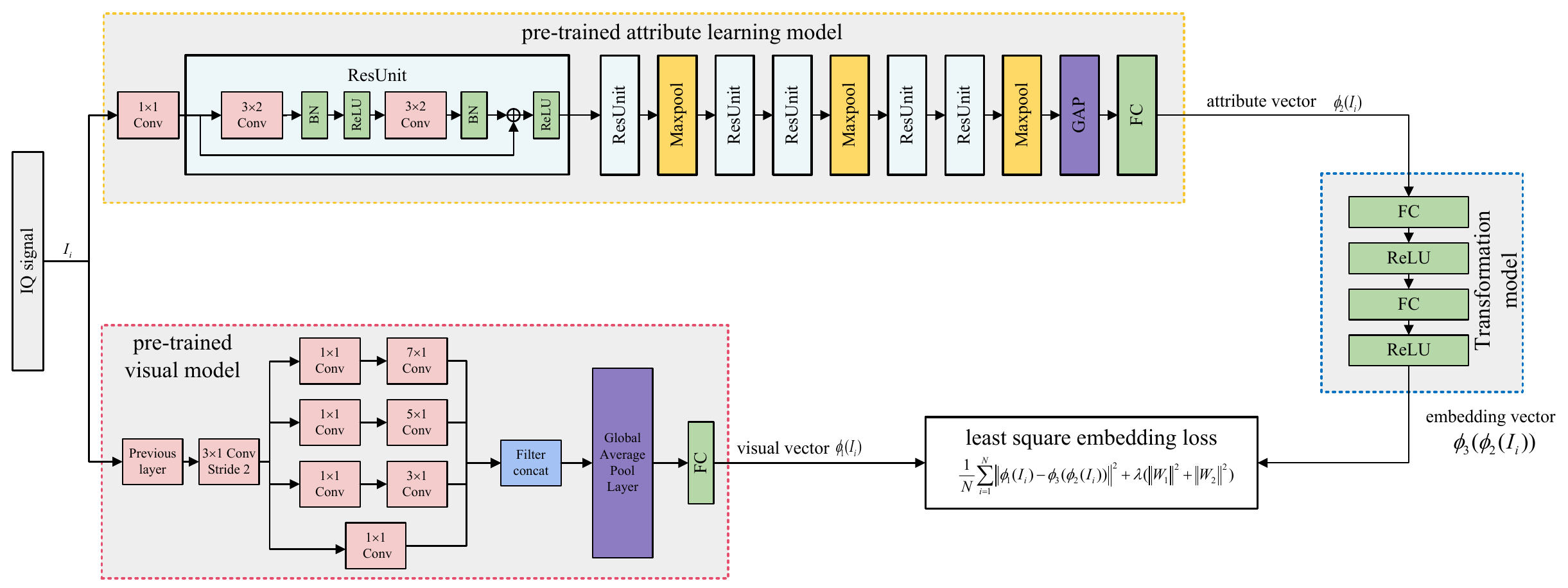}
\caption{The deep visual-semantic embedding model.}
\end{figure*}

Finally, the learned feature vectors are classified by another FC layer with the $softmax$ classifier. The number of neurons in the last layer is equal to the number of modulation formats. Thus, each output neuron corresponds to a modulation format. $Softmax$ is used to convert the output into the probability that the input signal belongs to each candidate modulation format. The cross-entropy loss function is utilized to measure the gap between the model output and the true label. It can be given as
\begin{align}
loss=\frac{1}{N}\sum\limits_{j=1}^{N}{\sum\limits_{k=1}^{K}{{{q}_{j,k}}\log ({{p}_{j,k}})}},
\end{align}
where $K$ represents the number of classes, $N$ represents the number of samples. ${p}_{j,k}$ is the output of $softmax$, which represents the probability when data sample $j$ belongs to class $k$. ${q}_{j,k}$ is a indicative variable, which is given as
\begin{align}
{{q}_{j,k}}=\left\{ \begin{array}{*{35}{l}}
   1,\text{if }j\text{ belongs to }k  \\
   0,\text{if }j\text{ does not belongs to }k.  \\
\end{array} \right.
\end{align}

However, the pure data-driven AMC method requires a large number of samples to complete the training of deep networks. Moreover, the classification accuracy of traditional methods is poor in the practical complex and dynamic scenarios, especially in the low SNRs. Fortunately, the exploitation of expert knowledge is promising to tackle those problems \cite{zhang2017learning}. However, to our best knowledge, few investigations have studied it. Therefore, we propose a novel data and knowledge dual-driven AMC method.


\section{Proposed AMC Scheme}\label{Proposed AMC Scheme}
The architecture of our proposed AMC scheme is shown in Fig. 1. On one branch, the visual model encodes the raw I/Q data into visual vectors. The visual feature space is used as the embedding space where both the visual content
and the attribute vector of the class that the modulation format belongs to are embedded.
On the other branch, the attribute learning model learns the attribute vectors corresponding to different modulation formats according to the deterministic attribute labels that are set artificially. Then, the transformation model converts the attribute vectors into the visual feature space. Lastly, a least square embedding loss is utilized to minimize the discrepancy between the visual feature vectors and the attribute vectors.

\subsection{Visual Model}
Fig. 2 illustrates the framework of the visual model. The multi-scale convolutional network (MSNet) is employed as the visual model. The MSNet consists of several multiscale (MS) modules, global average-pooling (GAP), FC layers and the $softmax$ classifier. MS modules are used to capture the multi-level feature information by using a $3\times 1$ convolution layer with stride = 2 to reduce the feature dimension at the top layer of the module. Then, several parallel convolutions with different kernel sizes are utilized to capture multi-level features. Features from different convolutional layers are consolidated together by using the concat operation. 
After the concat operation, the FC layer of the traditional CNN is replaced by the GAP for aggregating information from the MS modules to average each feature output from the four corresponding channels. Compared with the traditional FC layer, the over-fitting problem is avoided since there is no parameter required to be optimized in GAP. Moreover, another FC layer with rectified linear units (ReLU) is utilized to reduce the feature dimensionality. ReLU function can be expressed as
\begin{align}\label{27}\
f(z)=\max (0,z),
\end{align}
where $\max(\cdot)$ represents the operator for obtaining the maximum value.

The values of several neurons are set to be zero, which results in the sparsity of the network and reduces the interdependence among parameters. Therefore, the occurrence of over-fitting problems can be alleviated. Moreover, different from $sigmoid$ and $tanh$ activation function, ReLU function dose not have the saturation zone. Thus, the gradient vanishing problem can be avoided. The specific architecture of the visual model is shown in Table \ref{table1}.


\subsection{Attribute Learning Model}\label{subsectionB}
The core goal of AMC is to recognize the category of different modulations. Each modulation  has its unique characteristics.  Thus, modulation can be identified based on the high-level description that is phrased in terms of semantic attributes. Different classes are usually different in the high dimensional feature space. In this case, it is promising to combine semantic attributes and visual features to achieve a higher classification accuracy.

An adaptation of residual network was exploited to learn the attribute representation of different modulations. The residual unit is shown in Fig. 3(a). Two 2D convolution operations with kernel size $3\times3$ are performed in the residual unit. Different from ResNet in \cite{o2018over}, the batch normalization layer is used to standardize the data in the intermediate layers to avoid the gradient disappearance caused by the saturation of the partial derivative of the intermediate variable. Alongside, ReLU activation is used after the first convolution and after the skip connection to introduce non-linear operations as shown in Fig. 3(a). The residual stack, as shown in Fig. 3(b) consists of a convolution operation with a kernel size of $1\times1$ followed by two residual units. The maxpool layer is used to compress features. Moreover, the GAP is used to average the outputs of different channels. Therefore, the parameters are reduced greatly compared with FC in ResNet proposed in \cite{o2018over}. The specific architecture of the attribute learning network is shown in Fig. 3(c). The number of residual stack is decreased from 6 to 3 to reduce the complexity of the model.
The kernel size and output dimensions corresponding to different layers of the attribute learning model are shown in Table \ref{table2}.

Since the learning of each attribute transcends the specific classification task, the attribute learning model can be pre-learned independently \cite{zhang2017learning}. In this way, we can perform attribute learning through bigger datasets that are not limited to AMC datasets.

\begin{figure}[!t]
\centering
\includegraphics[width=3.4 in]{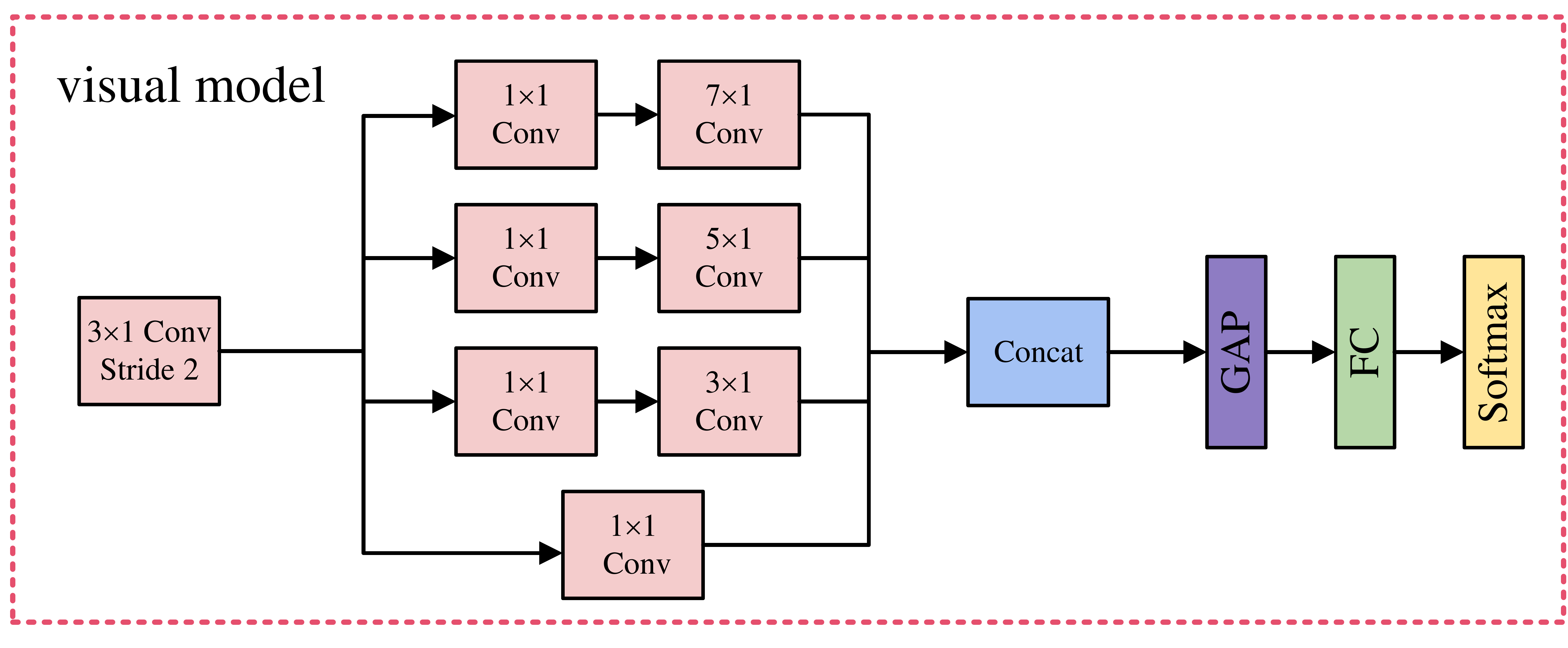}
\caption{The architecture of the visual model.} \label{fig.1}
\end{figure}

\begin{figure}
\centering
\subfigure[ ]{
\includegraphics[height=2.2in]{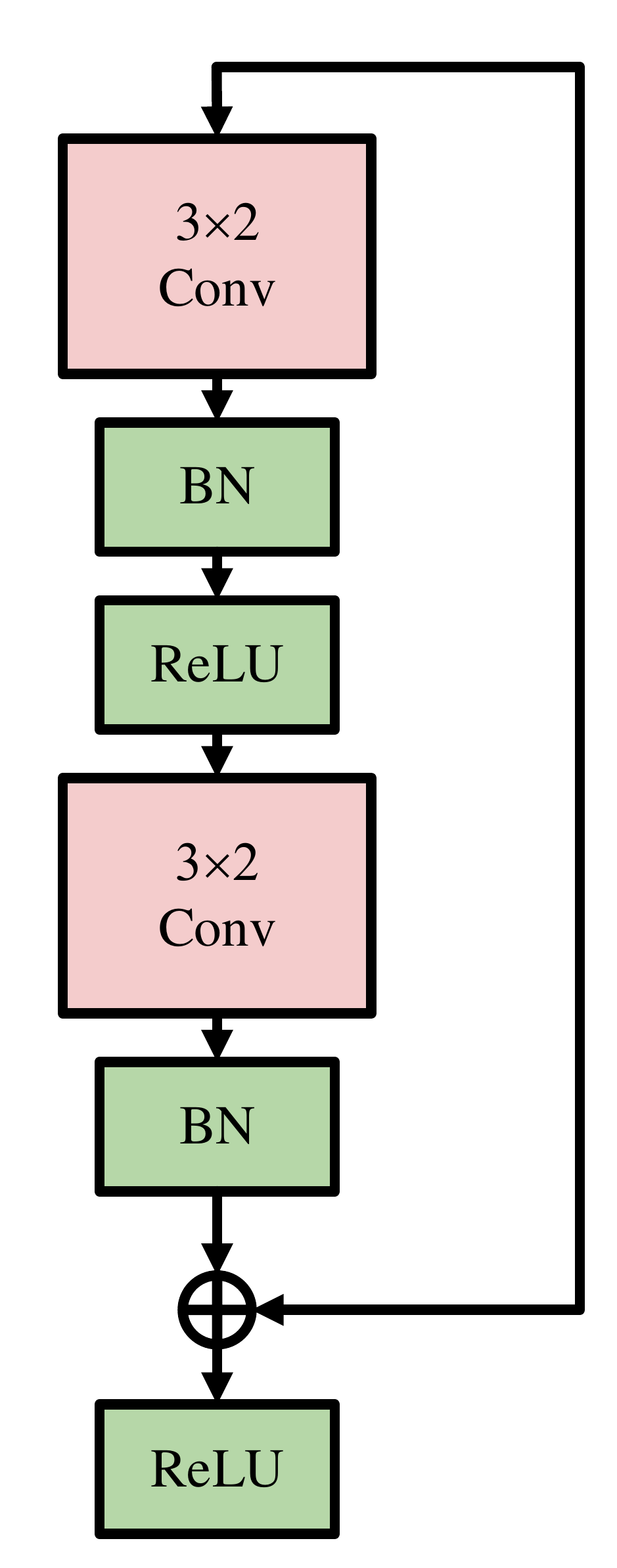}
\hspace{5mm}
}
\subfigure[ ]{
\includegraphics[height=2.2in]{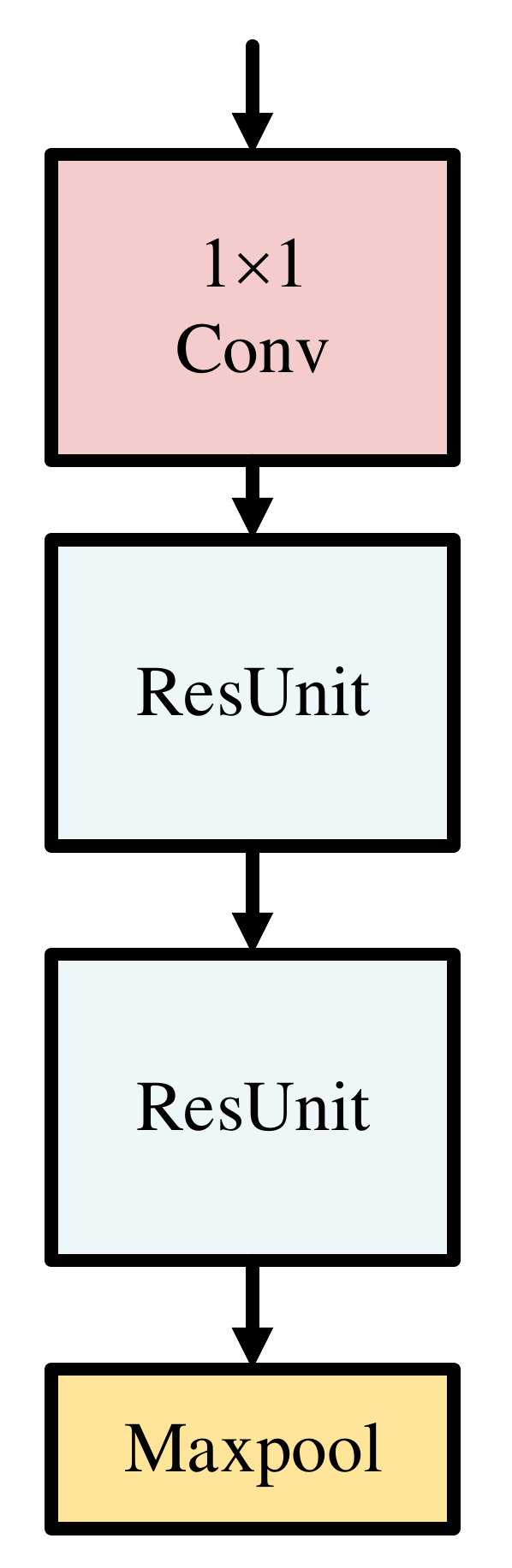}
\hspace{5mm}
}
\subfigure[ ]{
\includegraphics[height=2.2in]{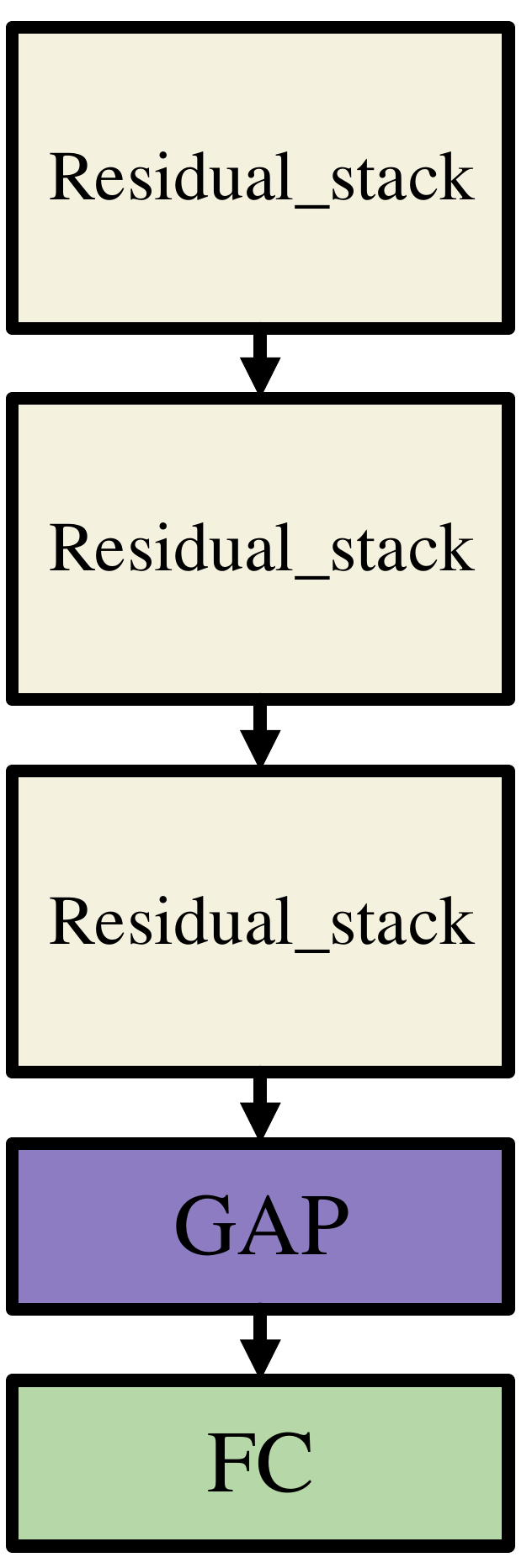}
\hspace{5mm}
}

\DeclareGraphicsExtensions.
\caption{(a) ResUnit. (b) Residual stack. (c) The attribute learning model.}
\end{figure}

\subsection{The Visual-Attribute Embedding Model}
The visual-attribute embedding model has two branches. One branch is the visual encoding branch adapted from the MSNet in Fig. 1, and the $softmax$ classification layer is removed. The visual encoding branch can output the visual feature vector. It takes the raw data of the signal samples $\mathbf{I}_{{i}}$ as the input. Then, the MS module is used to extract multi-dimensional features. The GAP is used to flatten multi-channel features. 
Finally, the FC layer outputs a $D$-dimensional feature vector $\phi_{1} ({\mathbf{I}_{i}}) \in {\mathbb{R}^{D\times 1}}$.

\begin{table}\normalsize\centering
\renewcommand\arraystretch{1.125}
 \caption{\label{tab:test}The Specific Architecture of the Visual Model}
\setlength{\tabcolsep}{0.5mm}
 \begin{tabular}{l|c|c}
  \midrule
  \midrule
  Input Layers & Kernel Size & Output \\
  \midrule
  \midrule
Input & $-$ & $2\times128$ \\
 MS module & $3\times1, [7\times1, 5\times1, 3\times1, 1\times1] $ & $128\times64$ \\
MS module & $3\times1, [7\times1, 5\times1, 3\times1, 1\times1] $ & $128\times32$ \\
GAP & $1\times1$ & $128\times4$ \\
FC/ReLU & $512\times128$ & $128$ \\
FC/Softmax & $128\times4$ & $4$ \\
\midrule
 \end{tabular}
 \label{table1}
\end{table}

\begin{table}\normalsize\centering
\renewcommand\arraystretch{1.125}
 \caption{\label{tab:test}The Specific Architecture of the Attribute Learning Model}
\setlength{\tabcolsep}{4mm}
 \begin{tabular}{l|c|c}
  \midrule
  \midrule
  Input Layers & Kernel Size & Output Dimensions \\
  \midrule
  \midrule
Input & $-$ & $2\times128$ \\
Conv & $1\times1$ & $32\times128$ \\
ResUnit & $3\times2$ & $32\times128$ \\
ResUnit & $3\times2$ & $32\times128$ \\
Maxpool & $2\times1$ & $32\times64$ \\

ResUnit & $3\times2$ & $32\times64$ \\
ResUnit & $3\times2$ & $32\times64$ \\
Maxpool & $2\times1$ & $32\times32$ \\

ResUnit & $3\times2$ & $32\times32$ \\
ResUnit & $3\times2$ & $32\times32$ \\
Maxpool & $2\times1$ & $32\times16$ \\

GAP & $1\times1$ & $32\times1$ \\
FC & $32\times6$ & $6$ \\

\midrule
 \end{tabular}
 \label{table2}
\end{table}
The semantic embedding is achieved by the other branch which is a attribute learning subnet as illustrated in \ref{subsectionB}. The subnet outputs a $L$-dimensional feature vector $\phi_{2} ({\mathbf{I}_{i}}) \in {\mathbb{R}^{L\times 1}}$. Moreover, a joint embedding space is learned where both the attribute vectors and the visual feature vectors can be projected. The authors in \cite{zhang2017learning} has proved that the visual feature space is the most appropriate embedding space, which can alleviate the hubness problem \cite{zhang2017learning}. Therefore, a transformation subnet is designed to convert attribute vectors into vectors in the same dimensional space as the visual feature vectors. Specifically, it takes the $L$-dimensional attribute representation vector $\phi_{2} ({\mathbf{I}_{i}}) \in {\mathbb{R}^{L\times 1}}$ of the attribute learning subnet as input, and after going through two FC layers and the ReLU activation outputs a D-dimensional semantic embedding vector ${{\phi }_{3}}({{\phi }_{2}}({\mathbf{I}_{i}}))\in {\mathbb{R}^{D\times 1}}$. In order to show the process of the transformation subnet  in detail, the output can be expressed as
\begin{align}\label{27}\
{{\phi }_{3}}({{\phi }_{2}}({\mathbf{I}_{i}}))={{f}_{2}}({\mathbf{W}_{2}}{{f}_{1}}({\mathbf{W}_{1}}{{\phi }_{2}}({\mathbf{I}_{i}}))),
\end{align}
where $\mathbf{W}_{1} \in {\mathbb{R}^{L\times M}}$ and $\mathbf{W}_{2} \in {\mathbb{R}^{M\times D}}$ are the weights to be learned in the first FC layer and the second FC layer. The rectified linear units $f(\cdot )$ is used as the activation function to introduce non-linearity into the network.

Each of the FC layer has a ${l}_2$ parameter regularization loss.
The two branches are linked together by a least square embedding loss which aims to minimise the discrepancy between the visual feature vector $\phi_{1} ({\mathbf{I}_{i}}) \in {\mathbb{R}^{D\times 1}}$ and its class representation embedding vector in the visual feature space. The loss function is given as
\begin{align}
 L({\mathbf{W}_{1}},{\mathbf{W}_{2}})=&\frac{1}{N}\sum\limits_{i=1}^{N}{{{\left\| {{\phi }_{1}}({\mathbf{I}_{i}})-{{f}_{2}}({\mathbf{W}_{2}}{{f}_{1}}({\mathbf{W}_{1}}{{\phi }_{2}}({\mathbf{I}_{i}}))) \right\|}^{2}}} \\ \notag
&+\lambda ({{\left\| {\mathbf{W}_{1}} \right\|}^{2}}+{{\left\| {\mathbf{W}_{2}} \right\|}^{2}}).
\end{align}
where $N$ is the number of training samples and $\lambda $ is the hyper-parameter weighting the strengths of the two
parameter regularization losses against the embedding loss.

As is shown in Fig. 3, different from the traditional pure data-driven architecture, our proposed scheme constructs a embedding space where both visual features and attribute features can be projected. The integration of attribute features can improve the performance of the model in low SNRs. Meanwhile, the existence of two pre-trained models makes end-to-end training only be required for the transformation model. Therefore, the training speed of our proposed scheme can be extremely high. Moreover, due to the introduction of attribute knowledge, we can reduce the requirements for visual model data. Lastly, different from abstract visual features, the attribute features labels are composed of deterministic binary variables which have clear physical implications.


\section{Simulation Results} \label{Simulation Results}
In this section, simulation results are presented to evaluate the performance of our proposed AMC scheme and compare with the benchmark schemes. The simulation settings are based on the works in \cite{9463441}. The performance is measured on a system equipped by a 3.00-GHz CPU, 16GB RAM, and a single NVIDIA GeForce GTX 1660SUPER GPU.

\begin{figure}[!t]
\centering
\includegraphics[width=3.65 in]{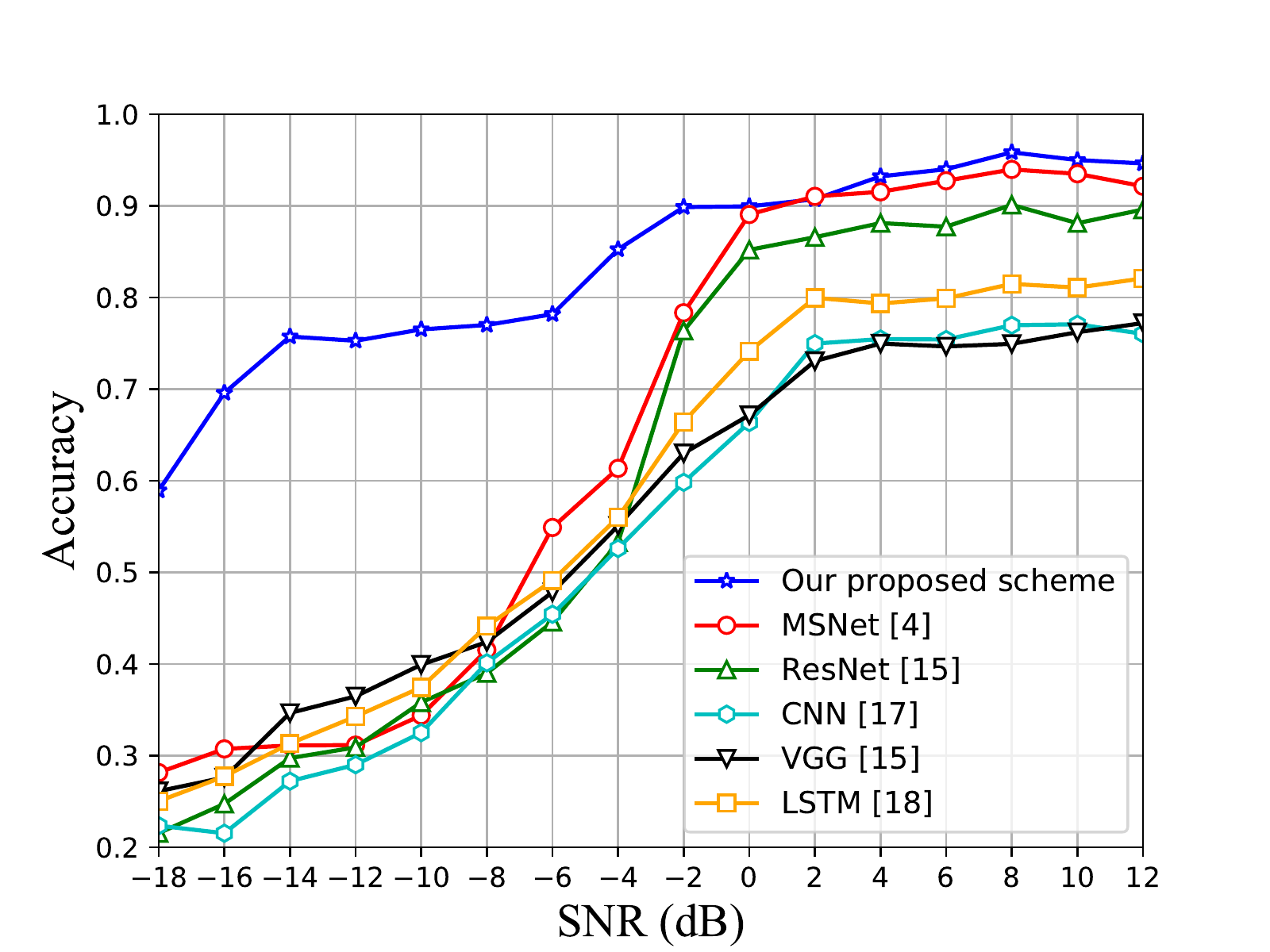}
\caption{Classification performance comparison among MSNet \cite{9463441}, ResNet \cite{o2018over}, CNN \cite{o2016convolutional}, VGG \cite{o2018over}, LSTM \cite{rajendran2018deep} and our proposed scheme.} \label{fig.1}
\end{figure}
\begin{figure}
\centering
\subfigure[ ]{
\includegraphics[height=1.65in]{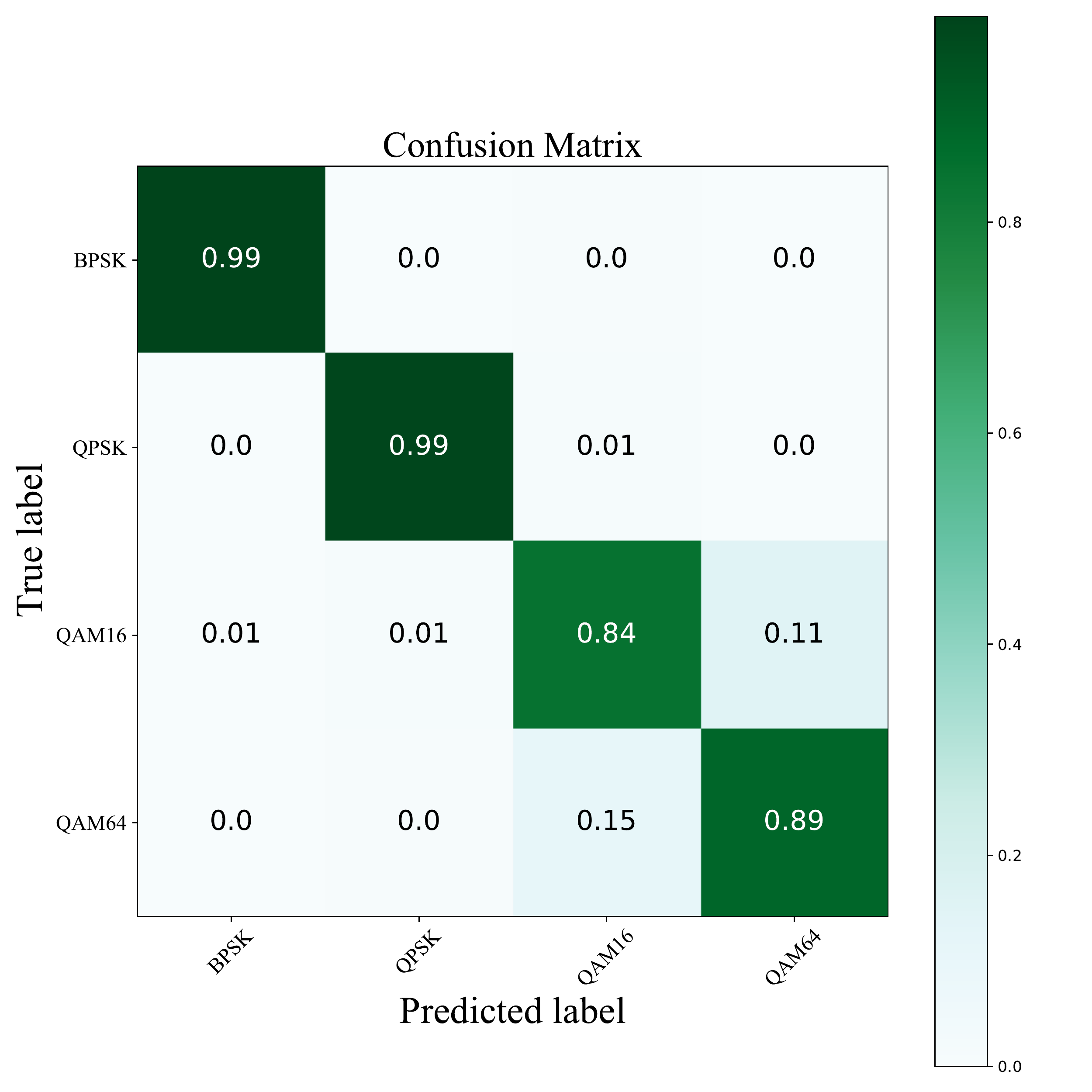}
}
\subfigure[ ]{
\includegraphics[height=1.65in]{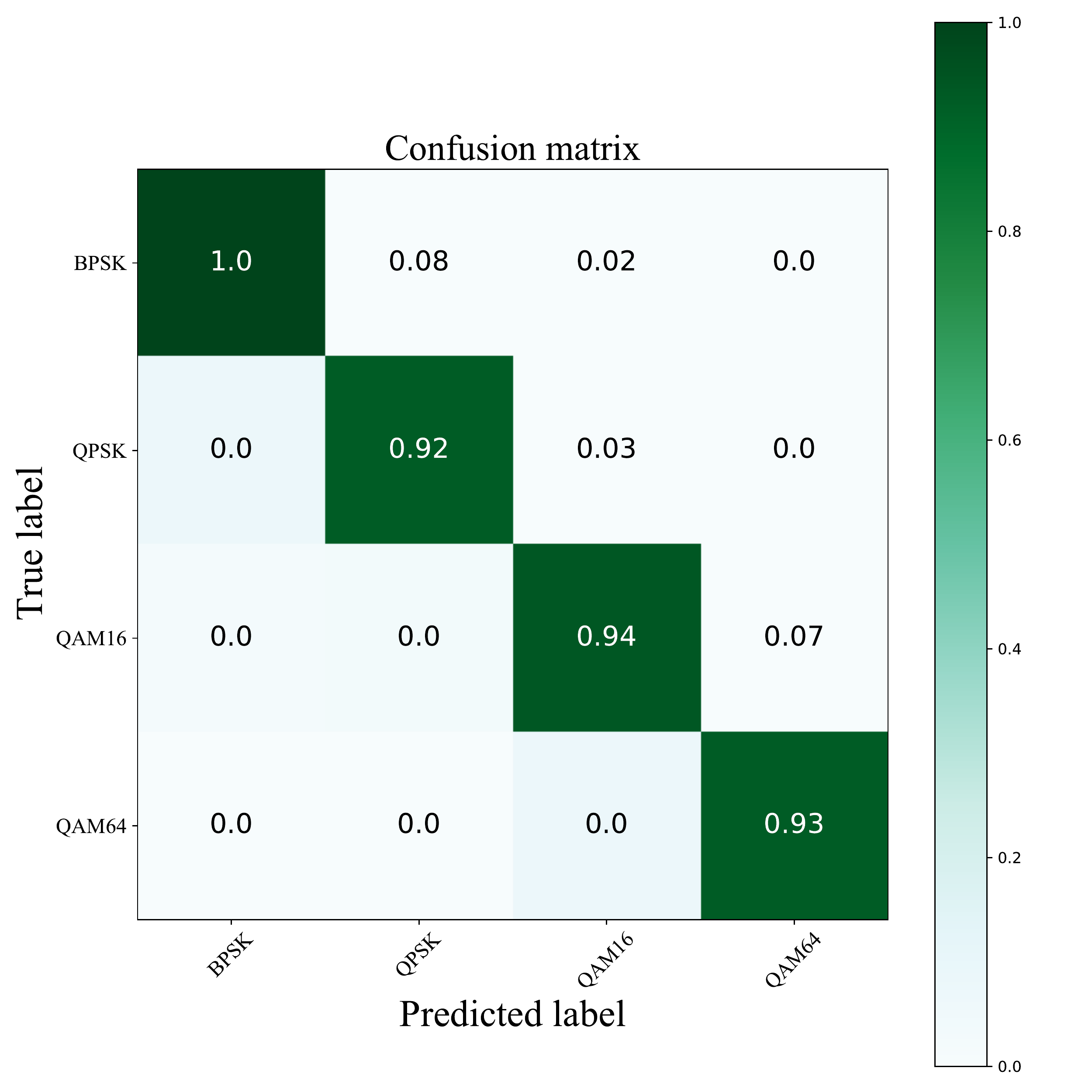}
}

\DeclareGraphicsExtensions.
\caption{Comparison of confusion matrices (a) MSNet \cite{9463441}. (b) Our proposed scheme. }
\end{figure}

\begin{figure}[!t]
\centering
\includegraphics[width=3.5 in]{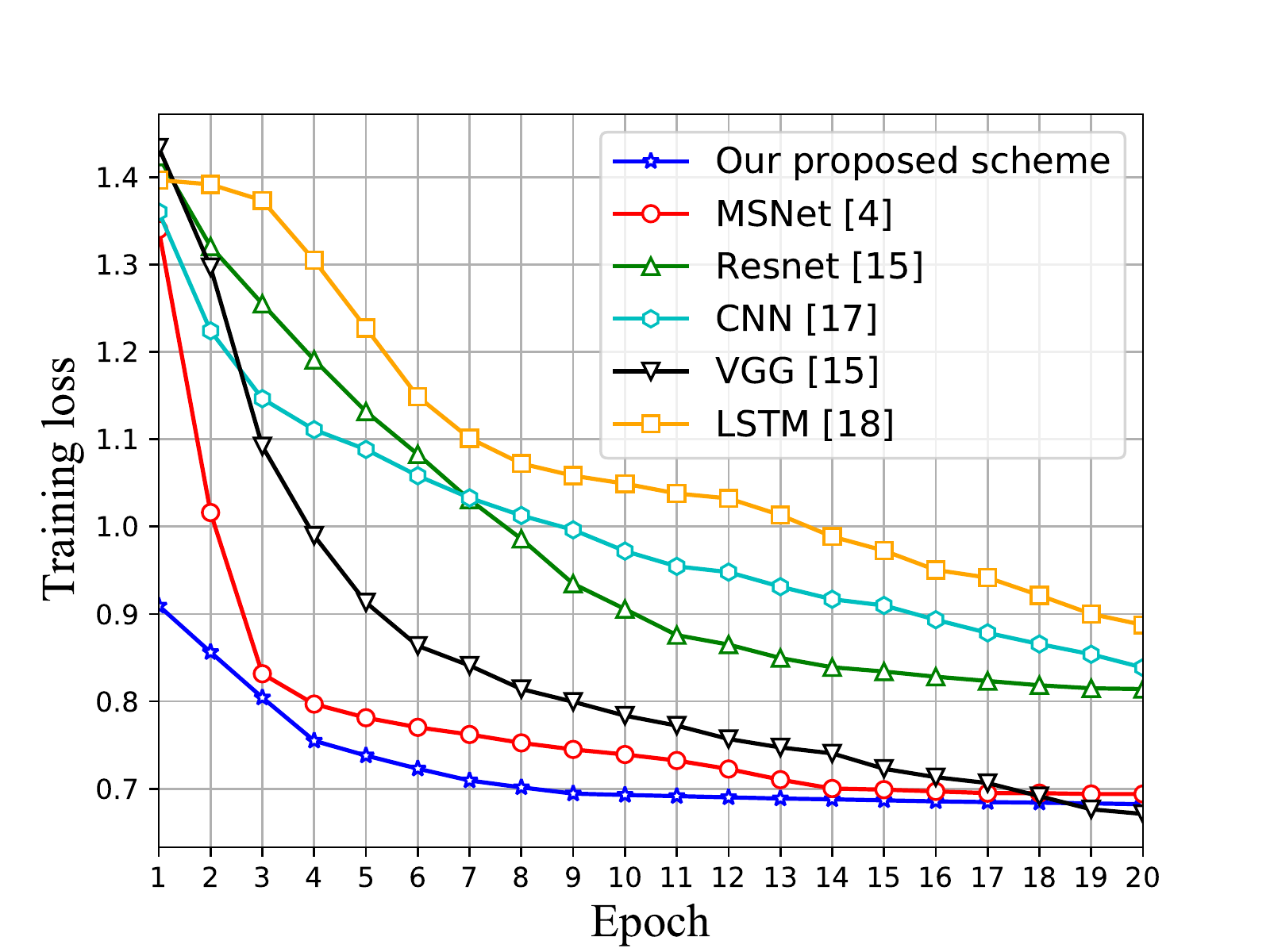}
\caption{Comparison of training loss between MSNet \cite{9463441}, ResNet \cite{o2018over}, CNN \cite{o2016convolutional}, VGG \cite{o2018over}, LSTM \cite{rajendran2018deep} and our proposed scheme.} \label{fig.1}
\end{figure}

\begin{figure}
\centering
\subfigure[ ]{
\includegraphics[height=1.9in]{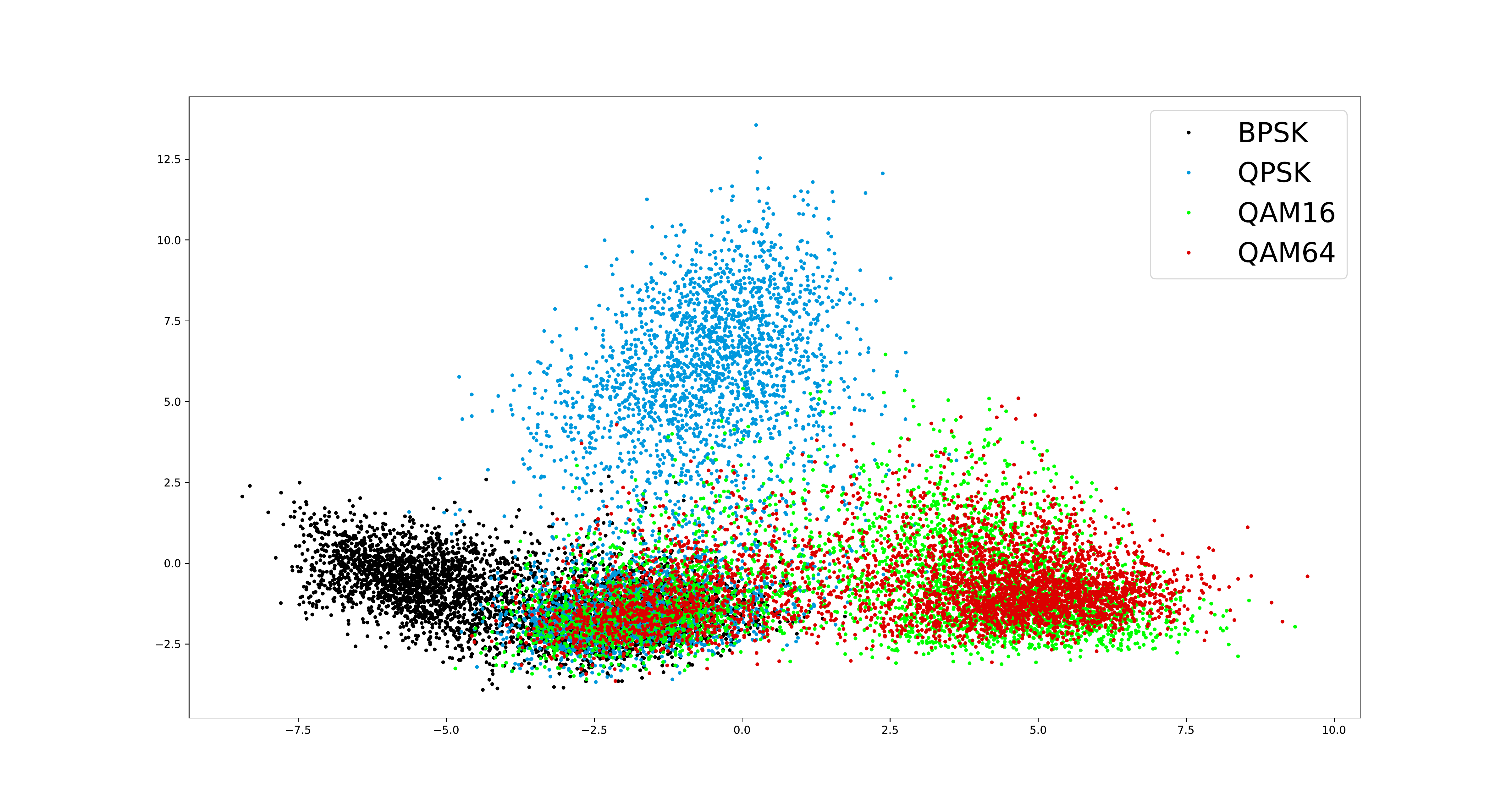}
}
\subfigure[ ]{
\includegraphics[height=1.9in]{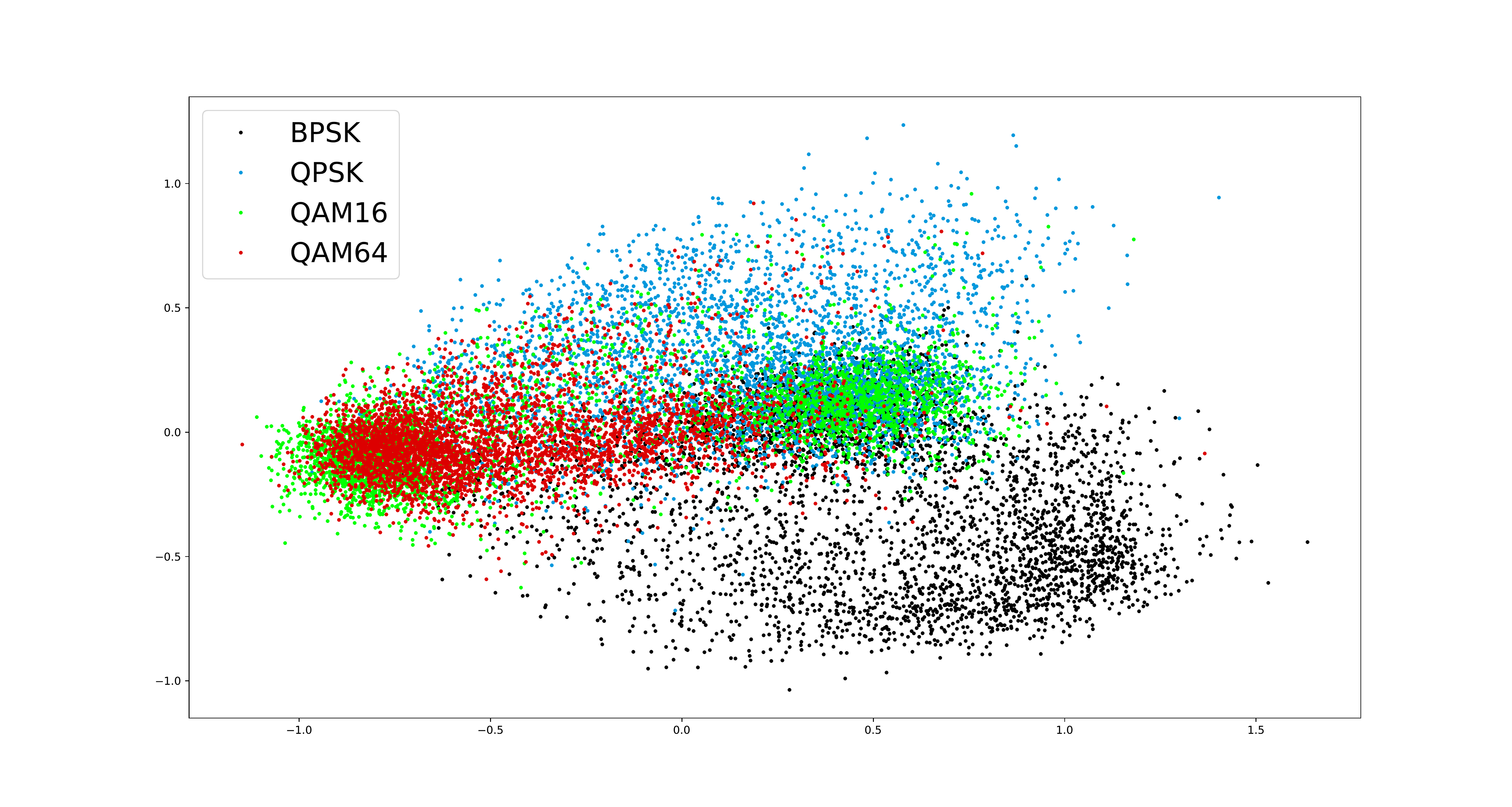}
}

\DeclareGraphicsExtensions.
\caption{Visualization scatter map of extracted features by different networks (a) MSNet \cite{9463441}. (b) Our proposed scheme. }
\end{figure}
For the attribute learning model, we train it by using the stochastic gradient descent (sgd) optimizer with an initial learning rate of 0.01, and a momentum of 0.9 for 40 epochs. Different from the training of visual model, the attribute learning is not a classification problem. Therefore, the mean squared error function is utilized to measure the gap between the model output and the true attribute label. 



The classification performance is operated on a publicly available dataset, which is presented on the website \footnote{https://www.deepsig.ai/datasets.}. Fig. 4 shows the classification performance comparison of our proposed scheme with those achieved by several representative DL-based models for AMC including MSNet \cite{9463441}, ResNet \cite{o2018over}, VGG \cite{o2018over} and LSTM \cite{rajendran2018deep}. It is evident that the proposed scheme is superior to other traditional models, and it can provide 2\% gains over ResNet, 5\% gains over MSNet, 17\% gains over LSTM and VGG at 12dB. Moreover, our proposed scheme has superior classification performance in low SNRs. It can reach about 60\% accuracy when the SNR is -20dB, and it can achieve over 70\% accuracy when the SNR is over -14dB, while the performance of ResNet is only about 25\% at -20dB and 32\% at -14dB.

The comparison of the confusion matrices between MSNet \cite{9463441} and our proposed scheme is shown in Fig. 5. It is seen that the proposed scheme has less confusion compared to other traditional models. Specifically, two modulation formats perform worse in MSNet, which are 16QAM and 64QAM as shown in Fig. 5(a) and Fig. 5(b), respectively.

Fig. 6 illustrates the training loss of each model for AMC. The proposed scheme can achieve a lower loss in the training set compared to the other models and obtains a higher convergence speed than other models. Specifically, our proposed model can achieve convergence at only about the $9$th epoch. This is reasonable, since the pre-trained subnets can greatly reduce the end-to-end training time.

To further demonstrate the effectiveness of the proposed scheme, we visualize the extracted feature and convert them into a two-dimensional scatter map \cite{wold1987principal}
shown in Fig. 7. It can be seen that 16QAM and 64QAM are mixed together in MSNet \cite{9463441}. In contrast, the features between 16QAM and 64QAM are much more discriminative in our proposed scheme. This indicates that our proposed scheme can achieve better classification performance for high-order modulations.
\section{Conclusion} \label{Conclusion}
A novel data and knowledge dual-driven AMC scheme based on RFML was proposed by exploiting attribute features and visual features. The attribute learning model was utilized to learn different attribute representations. The transformation model was used to convert attribute features into visual space to embed with the output of visual model. Simulation results demonstrated that our proposed scheme is superior to other benchmark schemes in terms of the classification accuracy especially in the low SNR. Moreover, the confusion between high-order modulations is reduced compared with other benchmark schemes.



\end{document}